\documentclass[twocolumn,aps,showpacs,prl]{revtex4}
\usepackage{graphicx}

\begin{document}
\title{Theoretical interpretation of Warburg's impedance in electrolytic cells }
\author{ G. Barbero$^{1,2}$}
\affiliation{
$^1$ Dipartimento di Scienza Applicata e Tecnologia, Politecnico di Torino,\\
Corso Duca degli Abruzzi 24, 10129 Torino, Italy.\\
$^2$ Moscow Engineering Physics Institute (MEPhI), National Research Nuclear
University, Kashirskoye shosse 31, 115409 Moscow, Russian Federation.\\
Correspondence to giovanni.barbero@polito.it}
\date{\today}

\begin{abstract}
We discuss the origin of Warburg's impedance in electrolytic cells containing only one group of positive and one group of negative ions. Our analysis is based on the Poisson-Nernst-Planck model, where the generation-recombination phenomenon is neglected. We show that to observe  Warburg's like impedance the diffusion coefficient of the positive ions has to differen from that of the negative one, and furthermore that the electrodes have to be not blocking. We assume that the non-blocking properties of the electrodes can be described by means of an Ohmic model, where the charge exchange between the cell and the external circuit is described by means of an electrode conductivity. For simplicity we consider a symmetric cell. However, our analysis can be easily generalized to more complicated situations, where the cell is not symmetric and the charge exchange is described by Chang-Jaffe model, or by a linearized version of Butler-Volmer equation. Our analysis allows to justify the expression for Warburg's impedance proposed previously by several groups, based on wrong assumptions.
\end{abstract}

 \maketitle

\section{Introduction}
One of the open problem in electrochemistry is the theoretical interpretation of Warburg's impedance. In his pioneer paper \cite{warburg} this impedance was interpreted as related to the ionic diffusion only. Analyses similar to that proposed by Warburg have been proposed more recently, following the same scheme by several searchers. Among others Warburg's impedance has been discussed in \cite{gerisher,buck0,buck1,buck2,bisquert,li,ju,moya,mac5,jrm1,jrm2,lai}. We have recently \cite{pccp,pre} criticized the analysis based only on the diffusion current \cite{gerisher,buck0,buck1,buck2,bisquert,li,ju,moya}, because in this case it is impossible to define the impedance of the cell in the standard manner, since the electric current is position dependent.  The apparent inconsistency between the existence of a current position dependent, and vanishing at the infinite in the case of a half space, and the definition of the impedance of the system as the ration between the applied potential and the current entering into the sample, has been source of contention and puzzlement to physicists and chemists since the original paper of Warburg \cite{jrm1,jrm2,lai}.
In the present paper, we will work out the details of the solution of the problem for the simplest case of two univalent ions, in the absence of generation-recombination, when the electrodes can be described by an Ohmic model. For this simple case the mathematics is not so complicated,  and one can easily see through it to the underlying physical meaning of what is going on. In particular,  it is possible to show that Warburg's impedance takes origin from the difference of the diffusion coefficients of the positive and negative ions, and from the non blocking character of the electrodes.

\section{Model}
We consider an electrolytic cell containing ions. Their bulk density, in thermodynamical equilibrium, is $n_0$. When the thermodynamical equilibrium is perturbed the actual bulk density for the positive and negative ions are $n_p$ and $n_m$. The ionic currents densities of positive and negative ions are  ${\bf {J_p}}=-D_p\,\nabla n_p+\mu_p\,n_p {\bf E}$, and ${\bf {J_m}}=-D_m\,\nabla n_p-\mu_m\,n_m {\bf E}$,
where $D_p$, $D_m$, and $\mu_p$, $\mu_m$ are the diffusion and mobility coefficients of the positive and negative ions, respectively. The actual electric field in the medium is related to the net charge density by Poisson's equation $\nabla\cdot {\bf E}=(q/\varepsilon) (n_p-n_m)$,
where $q$ is the electric charge of the ions, assumed monovalent, and $\varepsilon$ the dielectric constant of the liquid, free of ions. For the frequency range considered by us the electric field can be considered conservative and related to the electric potential by ${\bf E}=-\nabla V$.

The conservation of particles is described by the equations of continuity for the two type of ions, that in the absence of generation-recombination are $n_{p,t}=-\nabla \cdot {\bf {J_p}}$, and  $n_{m,t}=-\nabla \cdot {\bf{ J_m}}$, where we use the comma notation $f_{,x}=\partial f/\partial x$, $f_{,xx}=\partial^2 f/\partial x^2$ and so on. We limit our considerations to a sample in the shape of slab of thickness $d$, and assume the validity of Einstein-Smolucowski relation $\mu_p/D_p=\mu_m/D_m=q/(K_BT)$. The cartesian reference frame used for the description has the $z$-axis normal to the limiting surfaces, coinciding with the electrodes, at $z=\pm d/2$. We indicate by
$u_p=(n_p-n_0)/n_0$, $u_m=(n_m-n_0)/n_0$, and $u_v=q V/(K_BT)$,
the relative variations of the ionic bulk densities of positive and negative ions,  and the electric potential, expressed in $v_{th}=K_BT/q$, respectively. With these definitions, the fundamental equations of the model are
\begin{eqnarray}
\label{6} u_{p,t}&=&D_p(u_p+u_v)_{,zz},\\
\label{7}u_{m,t}&=&D_m(u_m-u_v)_{,zz},\\
\label{8}u_{v,zz}&=&-(u_p-u_m)/(2 \lambda^2).
\end{eqnarray}
Instead of $D_p$ and $D_m$ we use the quantities $D$ and $\Delta$ defined by $D_p=D/(1-\Delta)$ and $D_m=D/(1+\Delta)$
from which it follows that
$D=2 D_p D_m/(D_p+D_m)$ and $\Delta=(D_p-D_m)/(D_p+D_m)$.
Diffusion coefficient $D$ coincides with the ambipolar diffusion coefficient \cite{gb}. We define the unit of time
$t_u=\lambda^2/D$, where $\omega_D=D/\lambda^2$ is Debye's circular frequency related to the ambipolar diffusion. In the following we use the dimensionless units $\zeta=z/\lambda$ and $\tau=t/t_u$.
 Consequently $-M\leq \zeta \leq M$, where $M=d/(2 \lambda)$. In terms of the dimensionless parameters and coordinates the fundamental equations of the problems can be rewritten as
\begin{eqnarray}
\label{12}(1-\Delta)u_{p,\tau}&=&(u_p+u_v)_{,\zeta \zeta},\\
\label{13}(1+\Delta)u_{m,\tau}&=&(u_m-u_v)_{,\zeta \zeta},\\
\label{14}u_{v,\zeta\zeta}&=&-(u_p-u_m)/2,
\end{eqnarray}
forming a linear system of partial differential equations. In the following we will be interested in the determination of the electrical impedance of the cell under investigation, to investigate its  dependence on the circular frequency of the external applied potential difference. For this reason we limit the analysis to the case where the external power supply is such that
$V(\pm d/2,t)=\pm (V_0/2)\,\exp(i \omega t)=\pm(V_0/2)\,\exp(i \Omega \tau)$, where $\Omega=\omega/\omega_D$ is the dimensionless circular frequency expressed in unit of the ambipolar Debye's circular frequency. In terms of dimensionless quantities the boundary conditions on the reduced electric potential $u_v$ are
\begin{equation}
\label{15}u_v(\pm M,\tau)=\pm(u_0/2)\,\exp(i \Omega \tau).
\end{equation}
Since the system of Eq.s(\ref{12},\ref{13},\ref{14}) is linear the steady state solutions we are looking for have the functional form $[u_p,u_m,u_v](\zeta,\tau)=[\phi_p,\phi_m,\phi_v](\zeta)\,\exp(i \Omega \tau)$.
Substituting this ansatz into Eq.s(\ref{12},\ref{13},\ref{14}) we get the system of ordinary differential equations
\begin{eqnarray}
\label{17}i\Omega(1-\Delta)\phi_p&=&\phi_p''+\phi_v'',\\
\label{18}i\Omega(1+\Delta)\phi_m&=&\phi_m''-\phi_v'',\\
\label{19}\phi_v''&=&-(\phi_p-\phi_m)/2,
\end{eqnarray}
where the prime means a derivation with respect to $\zeta$. The boundary conditions of the problem are related to the presence of the external power supply, Eq.s(\ref{15}), and to the nature of the electrodes. In the following we assume that the electrodes are identical in all the aspects, and that the exchange of electric charge on them is described by Ohmic's model $J_p=\kappa_p E$, and  $J_m=-\kappa_m E$,
for all $\tau$, at $\zeta=\pm M$. As we have shown elsewhere \cite{ioannis}, Ohmic's boundary conditions are equivalent to Chang-Jaffe  boundary conditions. In the following instead of $\kappa_p$ and $\kappa_m$ we use the quantities $\kappa$ and $\delta$ defined by $\kappa_p=\kappa(1+\delta)$, and $\kappa_m=\kappa(1-\delta)$, from which it follows that
$\kappa=(\kappa_p+\kappa_m)/2$, and  $\delta=(\kappa_p-\kappa_m)(\kappa_p+\kappa_m)$.
In terms of the dimensionless quantities and coordinates, Ohmic's boundary conditions can be rewritten as
\begin{eqnarray}
\label{23}\phi_p'+[1-h(1+\delta)(1-\Delta)]\phi_v'&=&0,\\
\label{24}\phi_m'-[1-h(1-\delta)(1+\Delta)]\phi_v'&=&0,
\end{eqnarray}
where $h=\kappa/\kappa^*$, and $\kappa^*=q D n_0/(K_BT)$. The case of blocking electrodes is obtained when $h=0$. We observe that even if $\delta=0$, there is an anisotropy in (\ref{23},\ref{24}) when $\Delta\neq 0$.

From Eq.s(\ref{12},\ref{13},\ref{14}) we get
\begin{eqnarray}
\label{25}\phi_p''-\,\frac{1+i2 \Omega(1-\Delta)}{2}\,\phi_p+\frac{1}{2}\phi_m&=&0,\\
\label{26}\phi_m''-\,\frac{1+i2 \Omega(1+\Delta)}{2}\,\phi_m+\frac{1}{2}\phi_p&=&0,
\end{eqnarray}
whose solutions are
\begin{eqnarray}
\label{27}\phi_p(\zeta)&=&C_{pa}\sinh(\mu_a \zeta)+C_{pb}\sinh(\mu_b \zeta),\\
\label{28}\phi_m(\zeta)&=&k_a C_{pa}\sinh(\mu_a \zeta)+k_b C_{pb}\sinh(\mu_b \zeta),
\end{eqnarray}
where $C_{pa}$ and $C_{pb}$ are integration constants,
\begin{eqnarray}
\label{29}\mu_{a,b}=\sqrt{\frac{1+2 i \Omega\mp\sqrt{1-4 \Omega^2 \Delta^2}}{2}},
\end{eqnarray}
are the characteristics complex lengths, and
\begin{eqnarray}
\label{31}k_{a,b}=-2\left[\mu_{a,b}^2-\,\frac{1+i2 \Omega(1-\Delta)}{2}\right].
\end{eqnarray}
Consequently, the $\zeta$-part of the reduced electric potential is given by
\begin{eqnarray}
\label{33}\phi_v(\zeta)=&-&\left\{\frac{1-k_a}{2\mu_a^2}C_{pa}\sinh(\mu_a \zeta)+\frac{1-k_b}{2\mu_b^2}C_{pb}\sinh(\mu_b \zeta)\right\}\nonumber\\
&+&C_v \zeta,
\end{eqnarray}
where $C_v$ is another integration constant. The integration constants $C_{pa}$, $C_{pb}$ and $C_v$ have to be determined by means of the boundary conditions (\ref{15},\ref{23},\ref{24}).

\section{Impedance of the cell}
The total electric current density is given by
\begin{equation}
\label{34}{\bf j}=q({\bf J_p}-{\bf J_{m}})+\varepsilon{\bf E}_{,t},
\end{equation}
where the first contribution represents the conduction current, and the second the displacement current. In the slab geometry ${\bf j}$ has just $z$-component, that in terms of dimensionless quantities is
\begin{equation}
\label{35}j=j_0\,\left\{\frac{\phi_p'}{1-\Delta}-\frac{\phi_m'}{1+\Delta}+2\,\frac{1+i \Omega(1-\Delta^2)}{1-\Delta^2}\,\phi_v'\right\}e^{i \Omega \tau},
\end{equation}
where $j_0=q n_0 D/\lambda$. The current density  $j(\zeta,\tau)$ is such that $\partial j/\partial z=0$,
as it is easy to verify by means of Eq.s(\ref{17},\ref{18},\ref{19}). Substituting (\ref{27},\ref{28},\ref{33}) into (\ref{35}) we get
\begin{equation}
\label{37}j=2j_0\,C_v\,\,\frac{1+i \Omega (1-\Delta^2)}{1-\Delta^2}\,\,e^{i \Omega \tau},
\end{equation}
which is $\zeta$ independent, as expected.

The electric impedance of the cell, defined by $Z=\Delta V(t)/(j(t) S)$, where $S$ is the surface area of the electrodes and $\Delta V(t)=V(d/2,t)-V(-d/2,t)$ the difference of potential applied to the cell by means of the external power supply, is found to be
\begin{equation}
\label{38}Z=R_u\,\,\,\frac{u_0 (1-\Delta^2)}{C_v[1+i \Omega (1-\Delta^2)]},
\end{equation}
where $R_u=\lambda^3/(\varepsilon D S)$
is an intrinsic resistance  defined in terms of the surface of the cell and the physical parameters of the medium.

As stated above, the integration constants $C_{pa}$, $C_{pb}$ and $C_v$ have to be determined by means of the boundary conditions (\ref{15},\ref{23},\ref{24}). They will not be reported in the paper because their expressions are rather large. We will discuss before in general the predicted frequency dependencies of the real, $R$,  and imaginary, $X$, parts of the impedance, and in particular the parametric plot of $-X$ versus $R$, numerically obtained. After that, by means of reasonable approximation, we will show from where is coming the Warburg dependence of $X$ versus $R$, not correctly explained before \cite{warburg,gerisher,buck1,buck2,bisquert,li,ju,moya}.

The anisotropy in the surface conductivity, $\delta$ does not play an important role in the frequency dependence of $Z$, and in the following this parameter will be assumed $\delta=0$. The parameters playing a fundamental role in the existence of Warburg dependence are the anisotropy in the diffusion coefficient $\Delta =(D_p-D_m)/(D_p+D_m)$, and the surface conductivity $h=\kappa/\kappa^*$. Since $\kappa^*$ plays the role of characteristic intrinsic surface conductivity, we will limit our analysis to the case $h\sim 1$. For what concerns $\Delta$ we assume $0\leq \Delta \leq 1$. Of course $\Delta=0$  corresponds to $D_p=D_m$, and $\Delta=1$, to $D_p\gg D_m$. The value of $M=d/(2\lambda)$ is usually very large, and in our numerical calculation it is assumed $M=10^3$.

In Fig.1 we show $r=R/R_u$, a, and $x=X/R_u$, b, versus $\Omega=\omega/\omega_D$, where $\omega_D=D/\Lambda^2$ is the Debye's circular frequency related to the ambipolar diffusion, for $M=10^3$, $h=1$ and $\Delta=0.2,0.4,0.6,0.8$. For $\Delta\neq 0$ and $h\sim 1$ the spectrum of $r$ versus $\Omega$ shows the existence of two plateaux: one related to the free diffusion $r_f=2(1-\Delta^2)M$, and the other to the ambipolar diffusion $r_a=2 M$. The calculation of the limit for $\Omega\to 0$ of $r$, by means of the full expression of $C_v$ gives, in the limit of large $M$,
\begin{equation}
\label{40}r_0= 2\,\frac{h (M-1)+1}{h},
\end{equation}
that for $M\gg 1$ and $h M\gg 1$ is $h$ independent, and equal to $r_0\sim 2 M=r_a$. In the same framework the spectrum of $-x$ versus $\Omega$ presents two maxima, at the frequencies $\Omega_{\ell}=[\pi/(2 M)]^2$ and $\Omega_h=1/(1-\Delta^2)$ related to the ambipolar and free diffusion, respectively \cite{gb}. The parametric plot of $-x$ versus $r$ gives information on the Warburg's like impedance. It presents a circle in the high frequency region, whose radius is very close to $r_f$. Decreasing the circular frequency, it presents the typical Warburg dependence, and decreasing further $\Omega$, the dependence is again of circular type, and the circle ends at $r_a$ \cite{jamnik}.
In Fig. 1c we show the parametric plot of $-x$ versus $r$. Increasing $\Delta$ decreases the radius of the circle in the high frequency region, related to the free diffusion, as expected since $r_f=2(1-\Delta^2)M$. A similar analysis for $h$ ranging from $0.1$ to $2$, shows that the parametric plot is independent, in this range, of $h$. This parameters is important only if it is rather small, i.e. when $hM\sim1$, limit that it is not of interest in our analysis.
\begin{figure}[htbp]
\centering
\includegraphics[width=0.50\textwidth]{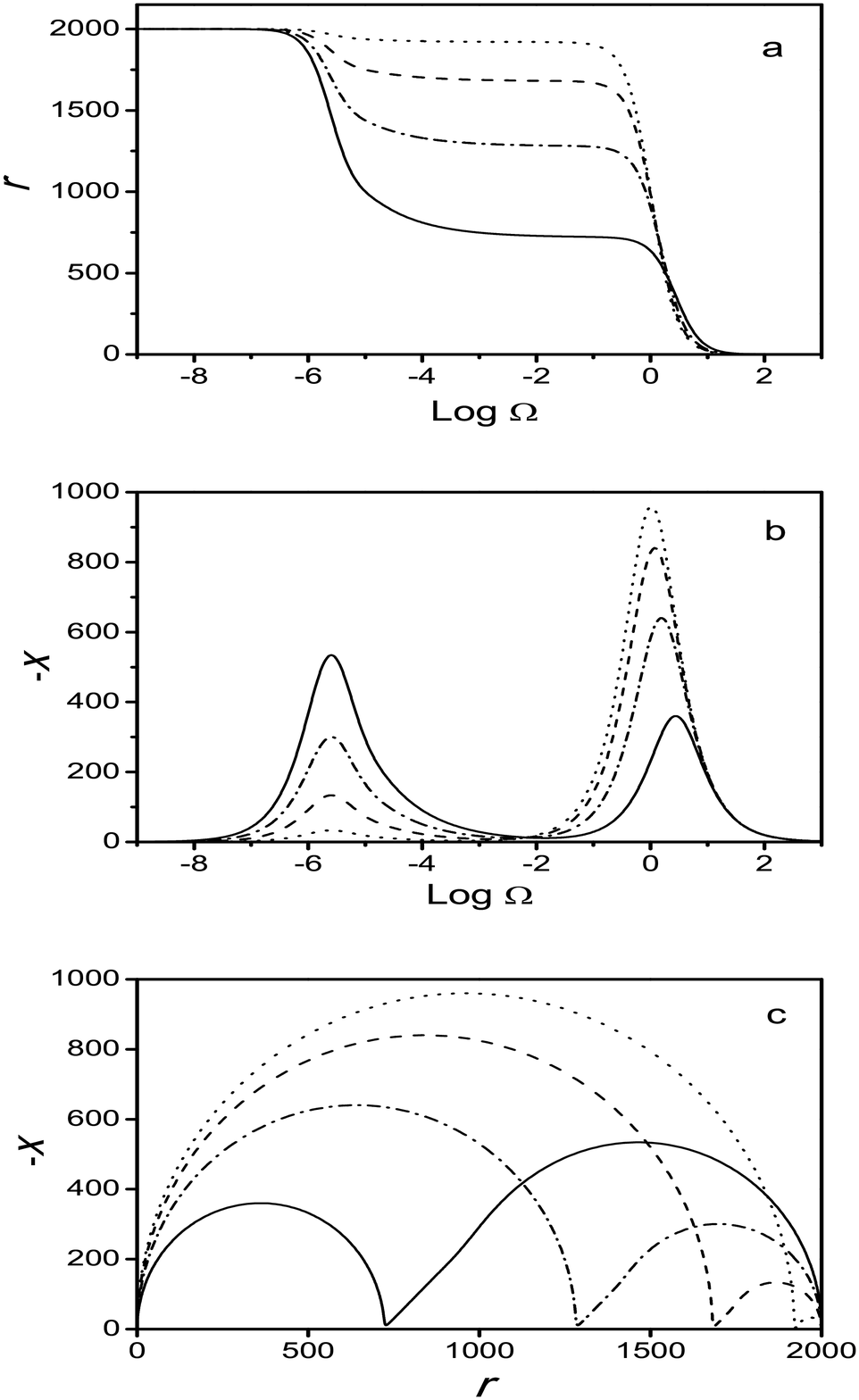}
\caption{Frequency dependence of $r=R/R_u$, a, $x=X/R_u$, b, and parametric plot of $-x$ versus $r$, c, for $h=1$, $M=10^3$ and $\Delta$ equal to $0.2$, dotted, $0.4$, dashed, $0.6$ dotted-dashed, and $0.8$, continuous. From c it is evident that increasing $\Delta$ the radius of the circle in the high frequency region decreases, and the linear part increases.}
\label{R}
\end{figure}
In the complete expression of $C_v$, and of $Z$, not reported, appear terms linear in $\Omega$ e terms of the type $1-(\Delta \Omega)^2$. Since the region of interest for Warburg's region corresponds to $\Omega<1$, and $\Delta<1$, we neglect $(\Delta \Omega)^2$ with respect to one. With this approximation the impedance of the cell, in the case of $\delta=0$, is given by
\begin{equation}
\label{41}Z=R_{\infty}\,\frac{(1-\Delta^2)+G+\Delta^2 \tanh(M\sqrt{i \Omega})/(M \sqrt{i \Omega})}{1+i(1-\Delta^2)\Omega},
\end{equation}
where $R_{\infty}=\lambda^2 d/(\varepsilon D S)$, and
\begin{equation}
\label{42}G=\frac{1-h-i\Delta^2}{M \sqrt{1+i \Omega}(h+i \Omega)}.
\end{equation}
Approximated formula (\ref{41}) coincides with the exact one in the whole frequency range.
From Eq.s(\ref{41},\ref{42}) it is possible to derive some rather important conclusions on the existence of Warburg's behaviour.
First of all we observe that for $\Omega\to 0$
\begin{equation}
\label{42-1}\Delta^2\,\frac{\tanh(M\sqrt{i \Omega})}{M \sqrt{i \Omega}}\to \Delta^2 ,\quad\quad G\to \frac{1-h- i \Delta^2}{M h},
\end{equation}
whereas for $\Omega \to \infty$ the two terms tend to zero, but
\begin{equation}
\label{42-2}\frac{\tanh(M\sqrt{i \Omega})}{M \sqrt{i \Omega}}\to \frac{1}{M \sqrt{i \Omega}} ,\quad\quad G\to \frac{1-h-i \Delta^2}{M (i \Omega)^{3/2}}.
\end{equation}
Hence, for $h\sim 1$ in the two limits the $G$ term is negligible with respect to the term $\tanh(M\sqrt{i \Omega})/(M \sqrt{i \Omega})$. A simple numerical analysis allows to verify that this is true in all frequency range when $h\sim 1$. Consequently Eq.(\ref{41}) is well approximated by
\begin{equation}
\label{41-1}Z=R_{\infty}\,\frac{(1-\Delta^2)+\Delta^2 \tanh(M\sqrt{i \Omega})/(M \sqrt{i \Omega})}{1+i(1-\Delta^2)\Omega},
\end{equation}
Formula (\ref{41-1}) for $\Delta \to 1$, i.e. for the case where one of the diffusion coefficient is very large with respect to the other,  can be rewritten as
\begin{equation}
\label{42-3}Z= R_{\infty} \frac{\tanh(M\sqrt{i \Omega})}{M\sqrt{i \Omega}},
\end{equation}
that coincides with the expression reported in \cite{warburg,gerisher,buck1,buck2,bisquert,li,ju,moya}. From this result it follows that, despite the analysis reported in  \cite{warburg,gerisher,buck1,buck2,bisquert,li,ju,moya} is not correct, the obtained result is sound. We stress  Eq.(\ref{42-2}) is valid only in the limit of $\Delta\to 1$, that means, f.i. $D_p\gg D_m$. In this case $D\sim 2 D_m$. Of course, for $\Delta=1$, i.e. $D_m=0$, $Z$ diverges, and Warburg's impedance is absent. We have already underlined that when only one group of ions is mobile, Warburg's impedance is not predicted by Poisson-Nernst-Planck model \cite{pccp,pre}.

In the pioneer paper of Warburg \cite{warburg}, where only the diffusion current was considered, the expression of the impedance is proportional to $\tanh(M\sqrt{i \Omega})/\sqrt{i \Omega}$.
This term is present in our analysis, and it is directly connected to $\Delta$. Hence, a condition to observe Warburg's impedance is $D_p\neq D_m$. However this condition is not enough. In fact, if the electrodes are blocking, and hence $h=0$, the $G$ term, defined by (\ref{42}), becomes
\begin{equation}
\label{44}G(h=0)=\frac{1-i\Delta^2}{i \Omega\,M\,\sqrt{1+i \Omega}},
\end{equation}
that, in the low frequency region, is more important of $\Delta^2 \tanh(M\sqrt{i \Omega})/(M \sqrt{i \Omega})$, and the linear term disappears.

Equation (\ref{41-1}) is more general than those reported in  \cite{warburg,gerisher,buck1,buck2,bisquert,li,ju,moya}, and can give information in the high frequency range. In fact, for $M\sqrt{\Omega}\gg 1$, that means $\Omega \gg 1/M^2$, $\tanh[M\sqrt{i \Omega}]=1$, Eq.(\ref{41-1}) can be rewritten as
\begin{equation}
\label{45}Z=R_{\infty}\frac{(1-\Delta^2)+(1-i) \Delta^2/(M \sqrt{2 \Omega})}{1+i(1-\Delta^2)\Omega},
\end{equation}
from which it is possible to derive the effective resistance and reactance, in the series representation, of the impedance of the  cell, along the lines suggested by \cite{jamnik,yeh}.

\section{Conclusions}
We have investigated the origin of Warburg's impedance in electrolytic cells. For simplicity we assumed that only one group of positive and negative ions are  present in the liquid and that the recombination-generation of ions can be neglected. The presented theoretical analysis is based on the Poisson-Nernst-Planck model, where the dynamical evolution of the bulk density of ions and the actual electric potential in the cell are described by the continuity equations and by the Poisson equation. The non blocking character of the electrodes is described by means of an Ohmic model. We have shown that to observe  Warburg's like impedance the diffusion coefficient of the positive ions has to differen from that of the negative one, and furthermore that the electrodes have to be not blocking. Our analysis can be easily generalized to take into account more groups of ions, of different boundary conditions for the  charge exchange between the cell and the external circuit. The result of our analysis allows to justify the expression for Warburg's impedance proposed previously by several groups, based on wrong assumptions.

{\bf Acknowledgment}
 This work was supported by the MEPhI Academic Excellence Project (agreement with the Ministry of Education and Science of the Russian Federation of August 27, 2013, project no. 02.a03.21.0005). Many thanks are due to Antonio Scarfone for useful discussions.

\end{document}